\documentclass[%
reprint,
aps,
prb,
]{revtex4-2}

\usepackage{hyperref}
\usepackage{dcolumn}
\usepackage{bm}
\usepackage{amsmath}
\usepackage{natbib}
\usepackage{booktabs}
\DeclareMathAlphabet{\mathbit}{OT1}{cmr}{bx}{it}

\usepackage{docmute}
\usepackage{color}
\usepackage{ulem}

\usepackage{graphicx}
\usepackage{braket}

\begin{document}

%
%

\title{Theory of pair density wave on a quasi-one-dimensional lattice in the Hubbard model}

\author{Soma Yoshida$^{1}$, Keiji Yada$^{1}$ and Yukio Tanaka$^{1}$}
\affiliation{%
  $^1$~Department of Applied Physics, Nagoya University, Nagoya 464--8603, Japan
}%

\begin{abstract}
In this study, we examine the superconducting instability of a quasi-one-dimensional 
lattice in the Hubbard model based on the random-phase approximation (RPA) and the fluctuation 
exchange (FLEX) approximation. 
We find that a spin-singlet pair density wave (PDW-singlet) 
with a center-of-mass momentum of 2$k_{F}$ can be stabilized 
when the one-dimensionality becomes prominent toward the perfect nesting of the Fermi surface. 
The obtained pair is a mixture of
even-frequency and odd-frequency singlet ones. 
The dominant even-frequency component does not have nodal lines on the Fermi surface. 
This PDW-singlet state is more favorable as compared to RPA when self-energy correction is introduced in the FLEX approximation.
\end{abstract}

\pacs{pacs}

\maketitle

\thispagestyle{empty}

\section{INTRODUCTION}
%
 
%

Unconventional superconductors have been studied extensively in condensed matter physics \cite{Ueda91}. 
Theoretical research on unconventional pairing in
strongly correlated superconductors was initiated
by the discovery of high-T$_{c}$ cuprate \cite{Scalapino1995}. 
It is known that 
antiferromagnetic spin fluctuation favors spin-singlet $d$-wave pairing in the Hubbard model in a two-dimensional square lattice
\cite{Scalapino1986d,Miyake1986}. 
Furthermore, several unconventional superconductors with $d$-wave spin-singlet pairing have been found.
However, it has been clarified that in a quasi-one-dimensional Hubbard model,
spin-singlet $d$-wave pairing becomes unstable because of the 
overlap of the nodal lines of the gap function and the Fermi surface. 
In this case, the odd-frequency spin-singlet 
$p$-wave pairing becomes dominant because 
the nodal lines of the gap function can avoid the Fermi surface \cite{Shigeta,Shigeta2,Yanagi,shigeta2013possible}. 

Although there have been several studies on odd-frequency gap functions 
in strongly correlated systems since the proposal by Berezinskii, 
the physical properties of odd-frequency gap functions have not been examined yet \cite{Berezinskii,Belitz1,Balatsky,Balatsky2,Coleman,Coleman2,Vojta,
Fuseya,Hotta,Shigeta,Shigeta2,Kusunose,Kusunose2}.
There is a fundamental difficulty in formulating a uniform odd-frequency gap function 
without the center-of-mass momentum \cite{Fominov}. 
Thus, it is necessary to consider a nonuniform Cooper 
pair for the realization of odd-frequency pairing. 
One such possible Cooper pair is the so-called pair density wave (PDW), where the center-of-mass momentum is of the order of the Fermi momentum \cite{Coleman,HoshinoPRL,HoshinoPRB2014,Hoshino2016}. 
It is interesting to study the stability of a PDW in a quasi-one-dimensional system from this viewpoint.

This state is distinct from the so-called Fulde-Ferrell-Larkin-Ovchinnikov (FFLO) state, where the center-of-mass momentum is significantly less 
compared with $k_{F}$. 
Research on PDWs started in the context of $\eta$ pairing \cite{CNYang}. 
At present, the PDW has become a hot topic in the research on stripe and pseudogap phases in high-$T_{C}$ cuprates \cite{Agterberg2008,Shirit2008,Berg_2009,Berg2009PRB,lee2014amperean,wang2015coexistence,mross2015spin,soto2015quasi,
hamidian2016,li2020eta,Chakraborty2021}.
The PDW singlet pairing state is one of the candidates for the pairing in the quasi-one-dimensional Hubbard model because the standard spin-singlet $d$-wave pairing becomes unstable. 
Thus, it is essential to check whether a PDW is 
possible in the quasi-one-dimensional Hubbard model. 


The PDW states are expected to be stable in quasi-one-dimensional systems for the following reasons.
The two electrons forming the Cooper pair in the PDW state have wavenumbers of $\bm{k}$ and $-\bm{k}+\bm{Q}$ with $\bm{Q}$ as the center-of-mass momentum. 
Because it is favorable that two electrons forming the Cooper pair exist near the Fermi level, they need to have the nesting vector of the system as the center-of-mass momentum.
In a quasi-one-dimensional system, the Fermi surface consists of two pieces, one on the left side and the other on the right side, so there is a nesting vector that overlaps two Fermi surfaces.
Then, it is expected that the PDW state will be stable by choosing such nesting vector as the center-of-mass momentum.
In the conventional superconducting state without the center-of-mass momentum, the wavenumbers $\bm{k}$ and $-\bm{k}$ of the two electrons forming the Cooper pair are related by the inversion symmetry of the system.
On the other hand, in the PDW state, such a relationship does not exist. 
Therefore, the even parity state and the odd parity state are mixed.
Then, based on the Fermi-Dirac statistics, even (odd) frequency component of the spin singlet gap function must be even (odd) function in the momentum space when we choose the origin at $\bm{k}=\bm{Q}/2$.  
Here, we suppose fully gapped superconductivity originated from repulsive interaction since the nodal superconducting state is unstable in the quasi-one-dimensional system \cite{Shigeta}.
Under this assumption, the fully gapped gap function has different signs on the right and left sides of the Fermi surface. 
In PDW state, the momentum dependence of even frequency component of the spin singlet gap function is consistent with such fully gapped momentum dependence if we choose the nesting vector as the center-of-mass momentum.
Thus, PDW state is also the possible candidate of the pairing symmetry in the quasi-one-dimensional system.

In this study, based on the standard random phase approximation (RPA), we solved the linearized Eliashberg equation by comparing six possible pairings: \\
(i) even-frequency spin-singlet even-parity (ESE), \\
(ii) even-frequency spin-triplet odd-parity (ETO), \\
(iii) odd-frequency spin-singlet odd-parity (OSO), \\
(iv) odd-frequency spin-triplet even-parity (OTE), \\
(v) 2$k_{F}$ spin-singlet pair density wave (PDW-singlet), and \\
(vi) 2$k_{F}$ spin-triplet pair density wave (PDW-triplet). 
The center-of-mass momentum of the Cooper pair is zero for (i) to (iv). 
Among these, the competitive dominant pairings are the ESE $d$-wave, OSO $p$-wave, and 2$k_{F}$ spin-singlet PDW. 
When one-dimensionality becomes prominent with the 
good nesting condition of the Fermi surface, 
the most dominant Cooper pair becomes a PDW-singlet pair. 
The present PDW-singlet state is supported by the nesting of the Fermi surface because the regions where a Cooper pair can be formed are enhanced. 
The resulting gap function is a mixture of the even- and odd-frequency components. 
Remarkably, the dominant even-frequency component displays a momentum dependence similar to that of the OSO $p$-wave component. Under the fluctuation exchange (FLEX) approximation, 
the OSO pairing becomes unstable owing to the self-energy effect, and 
the parameter regions where the PDW-singlet state is stable become wider. 

The rest of this paper is organized as follows. 
In Section II, we describe the model and the formulation of the RPA and 
FLEX approximation. In Section III, we present the results of the calculations based on the Eliashberg equation and illustrate the eigenvalue and momentum dependencies of the 
gap function. In Section IV, we summarize the results obtained.

\section{MODEL AND FORMULATION}

\begin{figure}[htbp]
  \parbox{0.64\linewidth}{\centering
  \includegraphics[width=\linewidth]{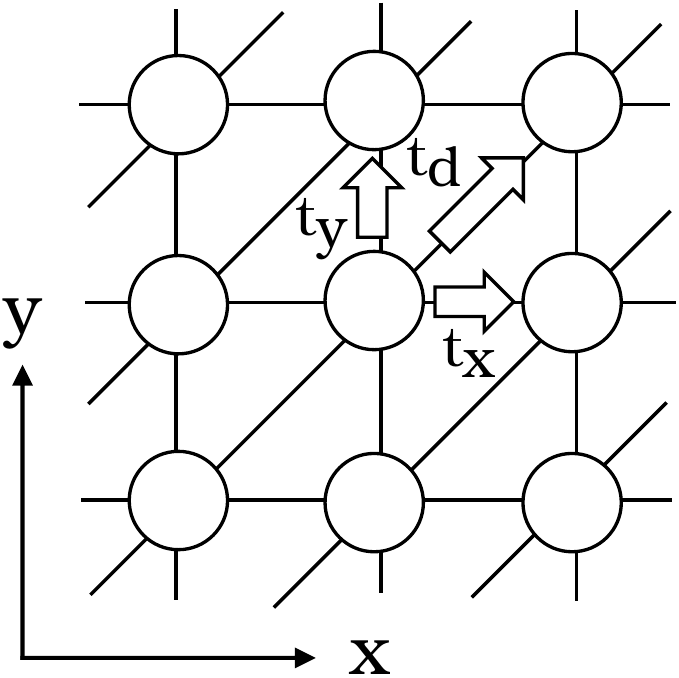}}
    \caption{Schematic of two-dimensional anisotropic Hubbard model}
\label{fig1}
\end{figure}

We analyze a two-dimensional single-band Hubbard model on an anisotropic triangular lattice, as depicted in Fig.~\ref{fig1}, where $t_x, t_y$, and $t_d$ are the hopping integrals along the $x$-, $y$-, and diagonal direction, respectively. The diagonal direction hopping makes the nesting of the Fermi surface incomplete, and helps to reveal the relation between the nesting and stability of the PDW state. Hereafter, we choose $t_x$ as the unit of energy.

\begin{figure}[htbp]
  \parbox{0.64\linewidth}{\centering
  \includegraphics[width=\linewidth]{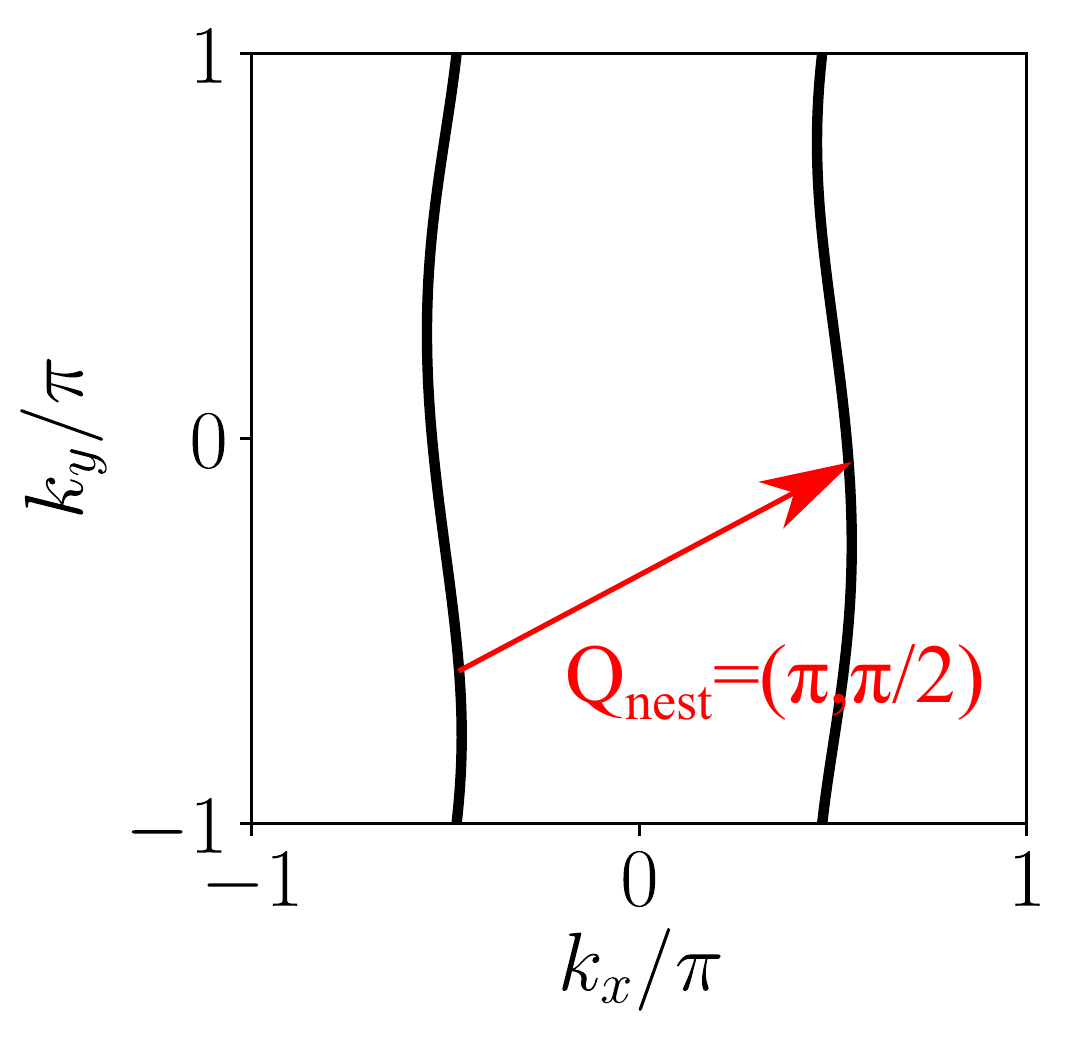}}
  \caption{Fermi surface in a quasi-one-dimensional region. 
  Nesting vector $\bm{Q}_{\mathrm{nest}}=(\pi,\pi/2)$. $t_y/t_x=t_d/t_x=0.1$.}
    \label{fig2}
\end{figure}

The Hamiltonian is expressed as\\
\begin{align}
  \hat{H}=\sum_{\braket{i,j},\sigma}\left(t_{ij}\hat{c}_{i\sigma}^\dagger\hat{c}_{j\sigma}+h.c.\right)+\sum_{i}U\hat{n}_{i\uparrow}\hat{n}_{i\downarrow},
\end{align}
where $t_{ij}$ is the hopping integral between the $i$ and $j$ sites, $\braket{i,j}$ represent the sets of nearest neighbors, $\hat{c}_{i\sigma}^\dagger(\hat{c}_{i\sigma})$ and $\hat{n}_{i\sigma}=\hat{c}_{i\sigma}^\dagger\hat{c}_{i\sigma}$ denote the creation (annihilation) and number operators, respectively, and
$U$ represents the on-site interaction. The dispersion relation in the normal state is represented as follows:\\
\begin{align}
  \varepsilon_{\bm{k}}=-2t_x\cos k_x-2t_y\cos k_y-2t_d\cos\left(k_x+k_y\right).
\end{align}
In this study, we consider the half-filling case, where the particle number $\hat{n}_i=\hat{n}_{i\uparrow}+\hat{n}_{i\downarrow}$ is equal to unity.
In the present model, the Fermi surface has a nesting vector $\bm{Q}_{\mathrm{nest}}=(\pi,\pi/2)$ in a quasi-one-dimensional region with $t_y/t_x=t_d/t_x\ll1$, as depicted in Fig.~\ref{fig2}.

In this study, we evaluate the superconducting pairing interactions based on the RPA \cite{Scalapino1986d,Miyake1986} 
and FLEX approximations \cite{Bulut,Bickers}.
\par Under these approximations, the 
irreducible susceptibility is expressed as follows:
\begin{align}
  \chi_0(q)=-\frac{T}{N}\sum_{k}G(k)G(k+q),\label{chi0} 
\end{align}
where $G(k)$ is the single-particle Green's function and $N$ is the number of $\bm{k}$-meshes.
Here, $k\equiv (\bm{k}, \varepsilon_n)$ and $q\equiv(\bm{q},\omega_m)$ are the short-hand notations of the momenta and Matsubara frequency for the fermion, $\varepsilon_n=(2n-1)\pi T$, and boson, $\omega_m=2m\pi T$, respectively.
Using the irreducible susceptibility, the spin susceptibility $\chi_{sp}(q)$ and charge susceptibility $\chi_{ch}(q)$ are expressed as follows:
\begin{align}
  \chi_{sp}(q)&=\frac{\chi_0(q)}{1-U\chi_0(q)}\label{chisp},\\
  \chi_{ch}(q)&=\frac{\chi_0(q)}{1+U\chi_0(q)}\label{chich}.
\end{align}
Here, the Stoner factor is defined as $f_s=U\chi_0(\bm{Q}_{\mathrm{nest}},0)$, as $\chi_0(q)$ is maximum at $q=(\bm{Q}_{\mathrm{nest}},0)$.
The magnetic phase transition occurs at $f_s\rightarrow1$, and the spin fluctuation becomes strong near the magnetic phase.
In this study, we choose the on-site interaction $U$ as $f_s=0.97$.
\par (a) Under RPA, the Green's function is expressed as $G(k)=\left(i\varepsilon_n-\varepsilon_{\bm{k}}+\mu\right)^{-1}$, where $\mu$ is the chemical potential and the self-energy is neglected.
\par (b) Under the FLEX approximation, Green's function is determined self-consistently using the Dyson equation as follows \cite{Bulut,Bickers}. 
First, we start from $G(k)=\left(i\varepsilon_n-\varepsilon_{\bm{k}}+\mu\right)^{-1}$ as Green's function. Next, we determine the spin and charge susceptibilities by eq.~$(\ref{chi0})$, $(\ref{chisp})$, and $(\ref{chich})$. Then, the effective interaction is calculated as follows:
\begin{align}
  V_n(q)=\frac{3}{2}U^2\chi_{sp}(q)+\frac{1}{2}U^2\chi_{ch}(q)-U^2\chi_0(q).
\end{align}
The self-energy is represented as follows:\\
\begin{align}
  \Sigma(k)=\frac{T}{N}\sum_{q}V_n(q)G(k-q).
\end{align}
Finally, we obtain the new Green's function as follows: \\
\begin{align}
  G^{-1}(k)=G_0^{-1}(k)-\Sigma(k).
\end{align}
We iterate these sequential calculations until Green's function converges sufficiently.
\par From the calculated spin and charge susceptibilities, we determine the effective pairing interaction for a spin-singlet and spin-triplet pairing as follows:
\begin{align}
  V_{\mathrm{eff}}^s(q)&=U+\frac{3}{2}U^2\chi_{sp}(q)-\frac{1}{2}U^2\chi_{ch}(q),\\
  V_{\mathrm{eff}}^t(q)&=-\frac{1}{2}U^2\chi_{sp}(q)-\frac{1}{2}U^2\chi_{ch}(q).
\end{align}
In the present study, we consider the possible realization of PDWs with the center-of-mass momentum $\bm{Q}$.
Therefore, the anomalous Green's function is defined as follows:
\begin{align}
F_{\sigma\sigma^\prime}(k,\bm{Q})=
-\int_0^\beta\langle T[\hat{c}_{\bm{k}\sigma}(\tau)\hat{c}_{-\bm{k}+\bm{Q}\sigma^\prime}(0)]\rangle e^{i\varepsilon_n\tau}d\tau.
\end{align}
This function must satisfy the following relation owing to Fermi--Dirac statistics: 
\begin{align}
F_{\sigma\sigma^\prime}(k,\bm{Q})=
-F_{\sigma^\prime\sigma}(-k+\bm{Q},\bm{Q}).\label{FDR}
\end{align}

The Dyson-Gorkov equation of the PDW state is noted as
\begin{align}
G(k)&=G_0(k)+G_0(k)
\Sigma(k)G(k)\nonumber
  \\&+G_0(k)\Delta_{\sigma\sigma^\prime}(k,\bm{Q})F^\dagger_{\sigma^\prime\sigma}(k,\bm{Q})\label{dge1}\\
F_{\sigma\sigma^\prime}(k,\bm{Q})&=G_0(k)\Sigma(k)F_{\sigma\sigma^\prime}(k,\bm{Q})\nonumber
                               \\&+G_0(k)\Delta_{\sigma\sigma^\prime}(k,\bm{Q})G(-k+Q)\label{dge2},
\end{align}
where we abbreviate $Q=(\bm{Q},0)$ and omit the summantion of the spin subscripts that appear repeatedly.
We linearize (\ref{dge1}) and (\ref{dge2}) with respect to the anomalous term.
Thus, the linearized Eliashberg equation is represented as
\begin{align}
  \lambda(\bm{Q})\Delta(k,\bm{Q})&=-\frac{T}{N}\sum_{k^\prime}V_{\mathrm{eff}}^{s,t}(k-k^\prime)F(k^\prime,\bm{Q}),\label{gapf}\\
  F(k^\prime,\bm{Q})&=G(k^\prime)G(-k^\prime+Q)\Delta(k^\prime,\bm{Q}),\label{F}
\end{align}
where we add the eigenvalue $\lambda(\bm{Q})$ to this equation to make it an eigenvalue equation. We calculate the maximum eigenvalue $\lambda$ and the gap function $\Delta(k)$ using the power method for possible pairings. 
It is known that the superconducting transition temperature 
is determined by the condition under which $\lambda$ becomes unity. 
Because $\lambda$ increases monotonically with decrease in temperature, 
the pairing with a larger value of $\lambda$ is more favorable.
\par 
In the PDW state, the inversion symmetry of the nomal state leads to the degenerate of the gap function $\Delta(k,\bm{Q})$ and $e^{i\theta}\Delta(k,-\bm{Q})$ as the solution of eq.~(\ref{gapf}) and (\ref{F}).
$\theta$ is an arbitrary phase and is not determined in the linearlized theory.
Thus, the inversion symmetry of the nomal state does not give any relation between $\Delta(k,\bm{Q})$ and $\Delta(-k+Q,\bm{Q})$.
Therefore, the even- and odd-parity states mix with each other. However, these anomalous terms do not break the spin-rotational symmetry of the system.
Thus, the pairing function is classified into the spin-singlet or spin-triplet states.
We refer to the singlet and triplet states that have a finite center-of-mass momentum as the PDW-singlet and PDW-triplet states, respectively.
From the Hermiticity of the Hamiltonian, the gap function as a solution to the linearized Eliashberg equation with a real eigenvalue satisfies $\Delta(\bm{k},\varepsilon_n)=\Delta^*(\bm{k},-\varepsilon_n)$. Therefore, we can choose the phase of the gap function, where the real (imaginary) part of the gap function has even (odd) in Matsubara frequency. Since the eigenvector of the Eiliashberg equation with eigenvalues does not necessarily satisfy this property, we require the eigenvectors to satisfy this property when we calculate the gap function numerically using the power method.
For example, the gap function of the PDW-singlet state satisfies
\begin{align}
  \mathrm{Real}[\Delta(\bm{k},\varepsilon_n,\bm{Q})]&=\mathrm{Real}[\Delta(-\bm{k}+\bm{Q},\varepsilon_n,\bm{Q})]\nonumber
                                                  \\&=\mathrm{Real}[\Delta(\bm{k},-\varepsilon_n,\bm{Q})],\label{FDR_sr}\\
  \mathrm{Imag}[\Delta(\bm{k},\varepsilon_n,\bm{Q})]&=-\mathrm{Imag}[\Delta(-\bm{k}+\bm{Q},\varepsilon_n,\bm{Q})]\nonumber
                                                  \\&=-\mathrm{Imag}[\Delta(\bm{k},-\varepsilon_n,\bm{Q})]\label{FDR_si}.
\end{align}
We take sufficiently large $N=N_x\times N_y$ $\bm{k}$-point meshes
with the cutoff Matsubara frequencies $\varepsilon_{M/2}$ and $\omega_{M/2}$.

\section{Results}

\par In the following, we study the PDW state in a quasi-one-dimensional parameter region, where the magnitudes of $t_y$ and $t_d$ are 
sufficiently lower than that of $t_x$.
\par
\begin{figure}[htbp]
	\parbox{0.78\linewidth}{\centering
  \includegraphics[width=\linewidth]{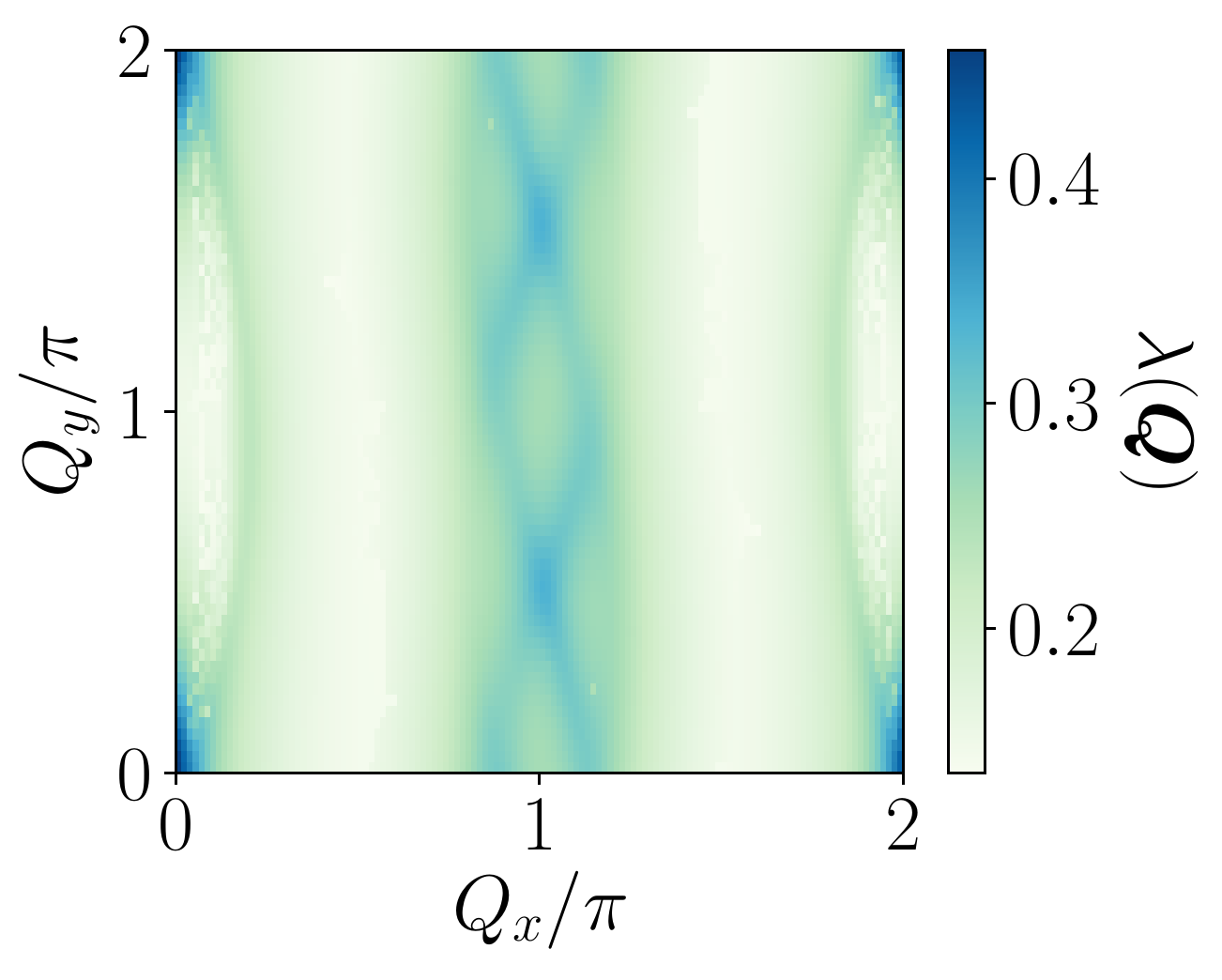}}
  \caption{Center-of-mass momentum dependencies of the maximum eigenvalue $\lambda(\bm{Q})$ of the linearized Eliashberg equation in the PDW-singlet state. $t_y/t_x=t_d/t_x=0.2$, $T/t_x=0.04$, $f_s=0.97$, $N=128\times64$, $M=512$.}
\label{fig3}
\end{figure}

\begin{figure}[htbp]
	\parbox{0.64\linewidth}{\centering
  \includegraphics[width=\linewidth]{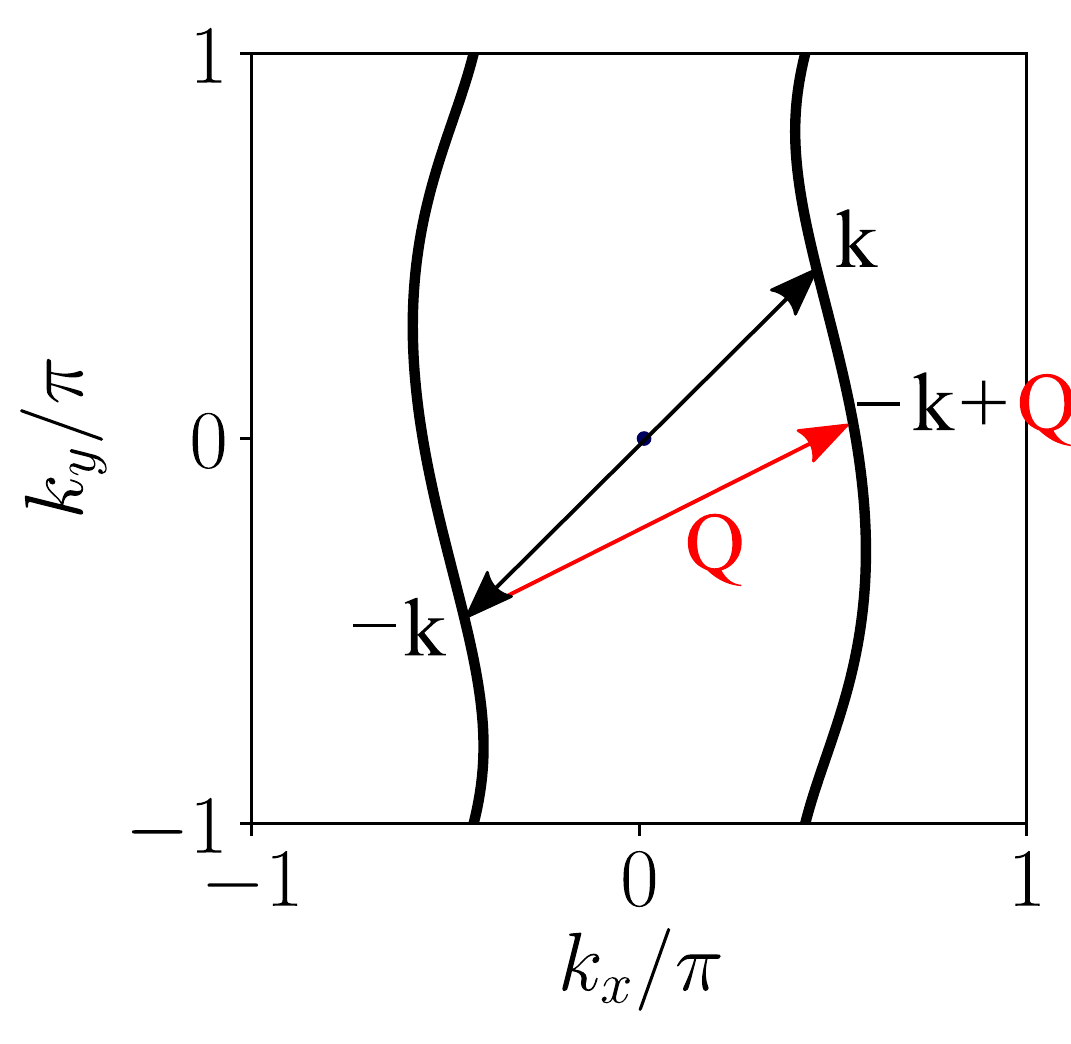}}
\caption{Two electrons that form a Cooper pair on the Fermi surface.}
\label{fig4}
\end{figure}

 We calculate the eigenvalues of the linearized Eliashberg equation 
by changing the center-of-mass momentum $\bm{Q}$ 
for $t_y/t_x=t_d/t_x=0.2$, as depicted in Fig.~\ref{fig3},
to obtain a value of $\bm{Q}$ that stabilizes the PDW state. 
In this parameter region, the maximum value of the eigenvalue are obtained as the center-of-mass momentum $\bm{Q}=\bm{0}$.
This means that a spatially uniform pairing without the center-of-mass momentum is 
obtained. In this case, the OSO pairing 
is obtained to be consistent with previous results \cite{Shigeta, Shigeta2}. 
On the other hand, the eigenvalue $\lambda(\bm{Q})$ has a peak at $\bm{Q}=\bm{Q}_{\mathrm{nest}}(=(\pi,\pi/2))$. 
Although the eigenvalue of $\bm{Q}=\bm{Q}_{\mathrm{nest}}(=(\pi,\pi/2))$ is lower than that of $\bm{Q}=\bm{0}$, it is expected that the solution with this nesting vector ${\bm Q}$ is realized by choosing smaller values of $t_y/t_x$ and $t_d/t_x$. 
This is explained by the fact that two electrons on the Fermi surface form a Cooper pair.
We suppose that two electrons have the wave numbers $\bm{k}$ and $-\bm{k}+\bm{Q}$.
As depicted in Fig.~\ref{fig4}, if an electron with $\bm{k}$ is located on the Fermi surface, then an electron with $-\bm{k}$ is also located on the Fermi surface owing to inversion symmetry. However, 
it is not evident that an electron with a wave number $-\bm{k}+\bm{Q}$ is located on the Fermi surface. 
To increase the number of electrons with momentum $-\bm{k}+\bm{Q}$
on the Fermi surface under the condition that an electron $\bm{k}$ is located on the Fermi surface,  
it is desirable that one side of the Fermi surface overlaps with the other side by the translation of momentum $\bm{Q}$.
Because this requirement agrees with the property of the nesting vector, 
it is appropriate to choose the nesting vector $\bm{Q}_{\mathrm{nest}}$ as the center-of-mass momentum of a Cooper pair.

\par
\begin{figure}[htbp]
	\parbox{0.96\linewidth}{\centering
  \includegraphics[width=\linewidth]{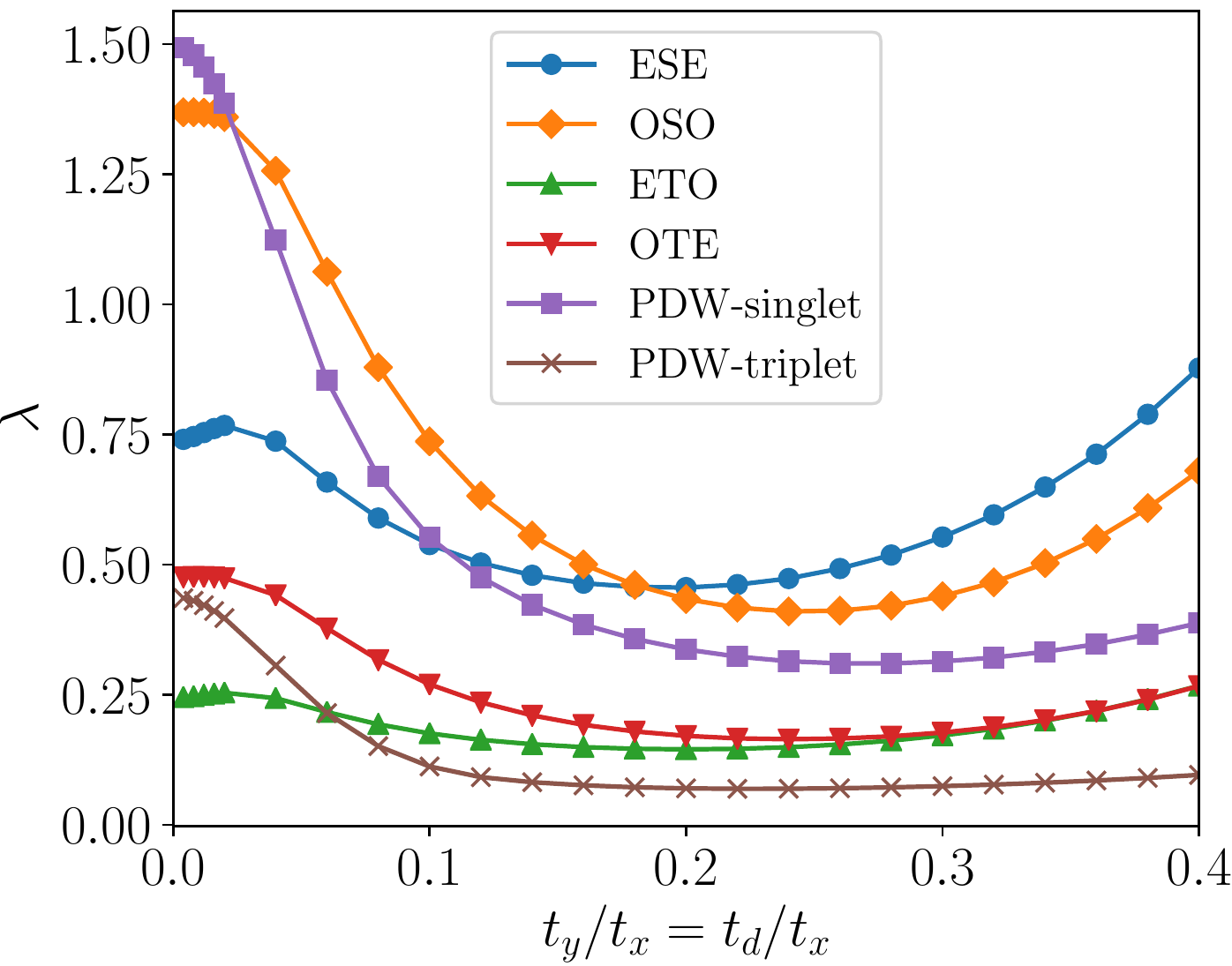}}
\caption{$t_y/t_x=t_d/t_x$ dependencies of the maximum eigenvalue $\lambda$ of the linearized Eliashberg equation. $T/t_x=0.04$, $f_s=0.97$, $N=128\times64$, $M=4096$}
\label{fig5}
\end{figure}

 After we obtain a preferable center-of-mass momentum, we study the stability of the PDW state in a quasi-one-dimensional system by changing $t_y/t_x=t_d/t_x$ and calculating the eigenvalues of each symmetry, as depicted in Fig.~\ref{fig5}, where we choose the nesting vector $\bm{Q}_{\mathrm{nest}}$ as the center-of-mass momentum $\bm{Q}$ of the PDW states. 
We obtain the nesting vector as the momentum that maximizes the irreducible susceptibility $\chi_0(\bm{q},0)$, which has a peak at the nesting vector~\cite{shigeta2013possible}.
The PDW-singlet state is stabilized at the extreme one-dimensional parameter region depicted in Fig\ref{fig5}.
This can be explained from the viewpoints of the nodes of the gap function and the degrees of  
nesting of the Fermi surface. 

\begin{figure*}[htbp]
  \includegraphics[width=\linewidth]{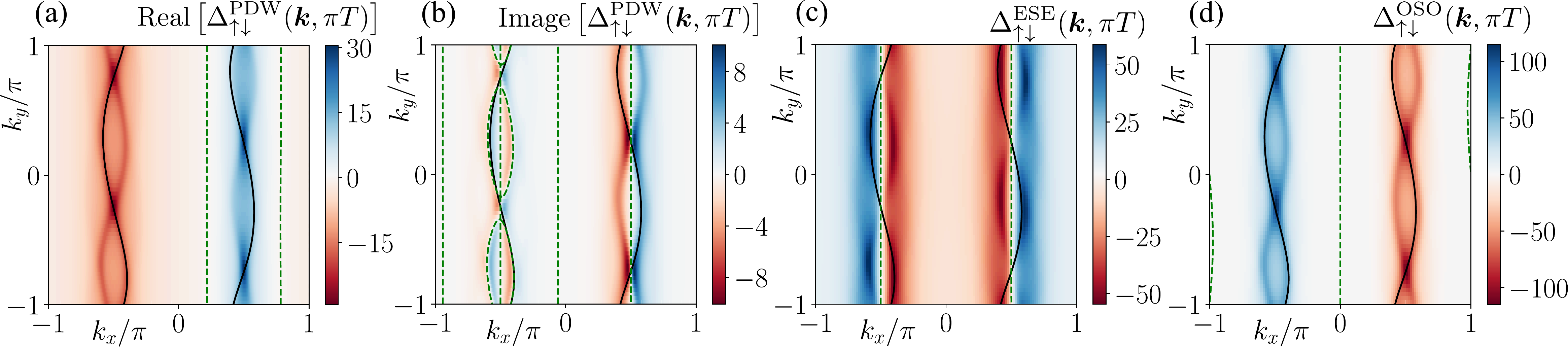}
  \caption{Momentum dependencies of (a) the real part of the  PDW state, (b) imaginary part of the PDW-singlet state, (c) ESE state and (d) OSO state gap functions. The solid lines represent the Fermi surface. Green dashed lines represent the node of the gap function. Center-of-mass momentum of the PDW-singlet state $\bm{Q}=(\pi,\pi/2)$, $t_y/t_z=t_d/t_x=0.2$, $T/t_x=0.04$, $f_s=0.97$,  $N=128\times64$, $M=4096$.}
    \label{fig6}
\end{figure*}

\begin{figure*}[htbp]
		\includegraphics[width=\linewidth]{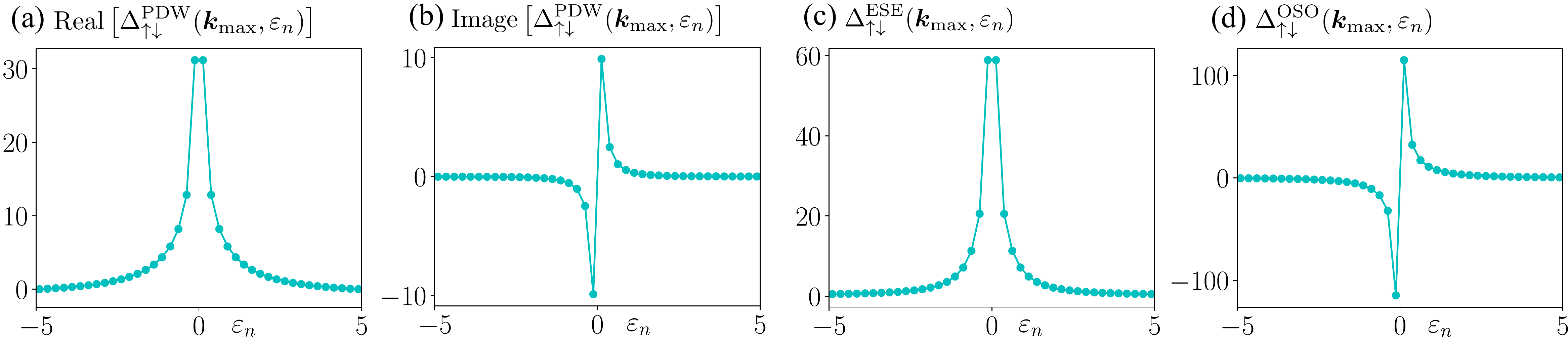}
    \caption{Matsubara frequency dependencies of (a) the real part of the  PDW state, (b) imaginary part of the PDW-singlet state, (c) ESE state and (d) OSO state gap functions. Center-of-mass momentum of the PDW-singlet state $\bm{Q}=(\pi,\pi/2)$, $t_y/t_z=t_d/t_x=0.2$, $T/t_x=0.04$, $f_s=0.97$,  $N=128\times64$, $M=4096$.}
    \label{fig7}
\end{figure*}

\par First, we illustrate the momentum and Matsubara frequency dependencies of the gap functions of the PDW-singlet, ESE, 
and OSO states in Figs.~\ref{fig6} and \ref{fig7}, respectively, to examine the stability of these pairings 
from the viewpoint of the number of nodes on the Fermi surface.
Figs.~\ref{fig6} show the momentum dependencies of $\Delta(\bm{k},\pi T)$. Figs.~\ref{fig7} show the Matsubara frequency dependencies of $\Delta(\bm{k}_{max},\varepsilon_n)$. $\bm{k}_{max}$ is the momentum that gives the maximum value of $|\Delta(\bm{k},\pi T)|$.
The solid and green dashed lines in Figs.~\ref{fig6} represent the Fermi surfaces of the normal state and nodes of the gap functions, respectively. 
The ESE and OSO states depicted in Figs.~\ref{fig6} and \ref{fig7} are consistent with the previous results
\cite{Shigeta2,shigeta2013possible}.
\par As referred to in the preceding study~\cite{Shigeta2,shigeta2013possible}, 
the OSO state is relatively stabilized because the ESE $d$-wave state is unstable in a quasi-one-dimensional parameter region owing to the overlapping of the nodes of the $d$-wave gap function and Fermi surface.
On the other hand, according to the momentum dependence of the PDW-singlet state in Fig.~\ref{fig6}(a), the line shape of the real part of the momentum dependence 
(even frequency part of the gap function) resembles that of the OSO state depicted in Fig.~\ref{fig6}(d) 
and does not have a node on the Fermi surface.
The corresponding line shape of the momentum dependence of 
the imaginary part (odd-frequency part of the gap function) of the PDW-singlet state in Fig.~\ref{fig6}(b) 
has four nodes on the Fermi surface, and the momentum dependence near $k_x=\pi/2$ is similar to that of the ESE state depicted in Fig.~\ref{fig6}(c). 
That is, the even- (odd-) frequency part of the PDW-singlet state appears to be odd (even) parity with respect to the origin. 
This reversal is caused by the generalized relation of the exchange of two electrons, as represented by eq.~(\ref{FDR}).
Eq.~(\ref{FDR_sr}) implies that the real part of the PDW-singlet state satisfies the equation 
\begin{align}
&{\rm  Real}
[\Delta_{\uparrow\downarrow}(\bm{k}+\bm{Q}/2,\varepsilon_n,\bm{Q})]\nonumber\\
&={\rm Real}[\Delta_{\uparrow\downarrow}(-\bm{k}+\bm{Q}/2,\varepsilon_n,\bm{Q})].
\end{align}
In contrast, the corresponding imaginary part satisfies the equation
\begin{align}
&{\rm  Imag}
[\Delta_{\uparrow\downarrow}(\bm{k}+\bm{Q}/2,\varepsilon_n,\bm{Q})]\nonumber\\
&=-{\rm Imag}[\Delta_{\uparrow\downarrow}(-\bm{k}+\bm{Q}/2,\varepsilon_n,\bm{Q})].
\end{align}
This means that the even-frequency (odd-frequency) part of 
the PDW-singlet state has an inversion symmetry with even(odd) parity with respect to ${\bm k}=\bm{Q}/2=(\pi/2,\pi/4)$. 
This property allows the even-frequency component
to have a momentum dependence similar to that of the OSO state when $\bm{Q}$ is a nesting vector.
Noted that even (odd) frequency part of the gap function of the PDW-singlet state is not strictly odd (even) function with respect to an inversion of $\bm{k}$ to $-\bm{k}$ and it is accidental that even frequency part has the momentum dependence similar to OSO p-wave.

The obtained ratio of the norms of the real and imaginary parts of the PDW-singlet state $\sum_{\bm{k},\varepsilon_n}|\mathrm{Real}[\Delta(\bm{k},\varepsilon_n)]|^2:\sum_{\bm{k},\varepsilon_n}|\mathrm{Imag}[\Delta(\bm{k},\varepsilon_n)]|^2$ at $t_y/t_x=t_d/t_x=0.2$ is equal to $166:1$. 
For this reason, the real part is dominant in the PDW-singlet state.
It is expected that the PDW-singlet state is stable because there are no nodes in the momentum and Matsubara frequency spaces. 

\begin{figure*}[htbp]
  \parbox{0.3\linewidth}{\centering(a)
		\includegraphics[width=\linewidth]{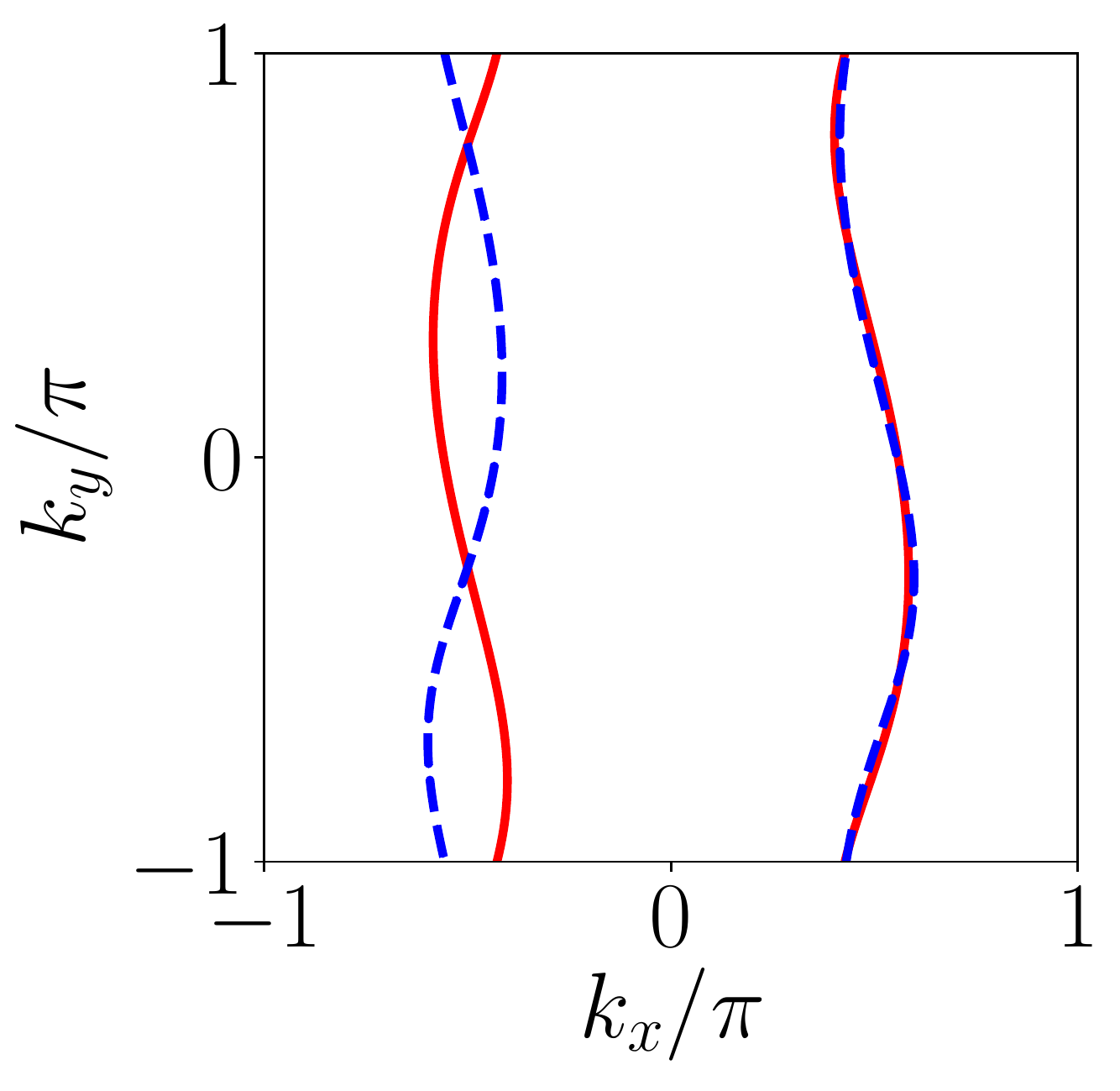}}
    \parbox{0.3\linewidth}{\centering(b)
  \includegraphics[width=\linewidth]{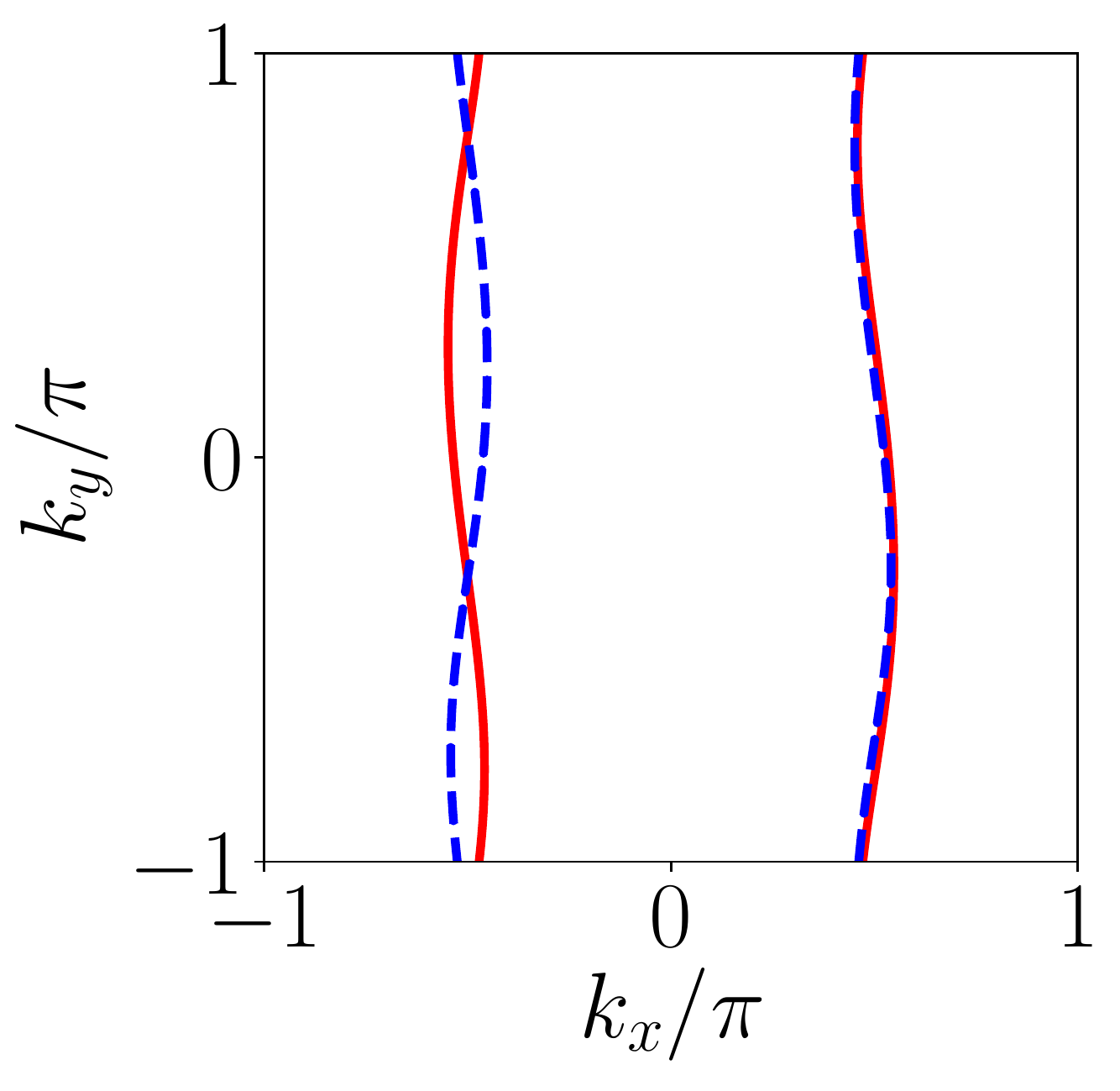}}
  \caption{Nesting of the Fermi surface. (a)$t_y/t_x=t_d/t_x=0.2$, (b)$t_y/t_x=t_d/t_x=0.1$. The red solid and blue dashed lines represent the Fermi surface and one that is parallelly moved by $\bm{Q}=(\pi,\pi/2)$, respectively.}
  \label{fig8}
  \parbox{0.4\linewidth}{\centering(a)
  \includegraphics[width=\linewidth]{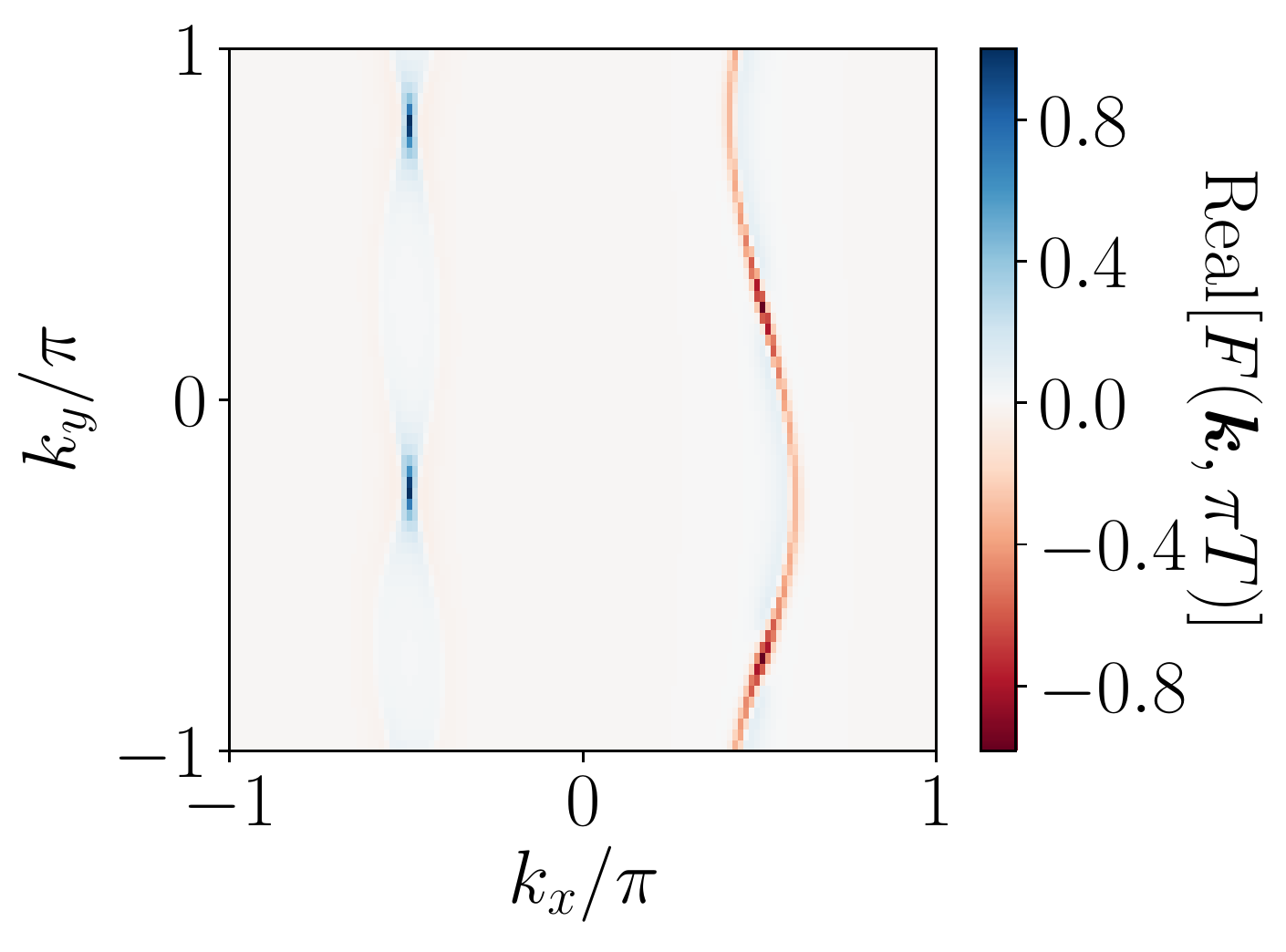}}
  \parbox{0.4\linewidth}{\centering(b)
  \includegraphics[width=\linewidth]{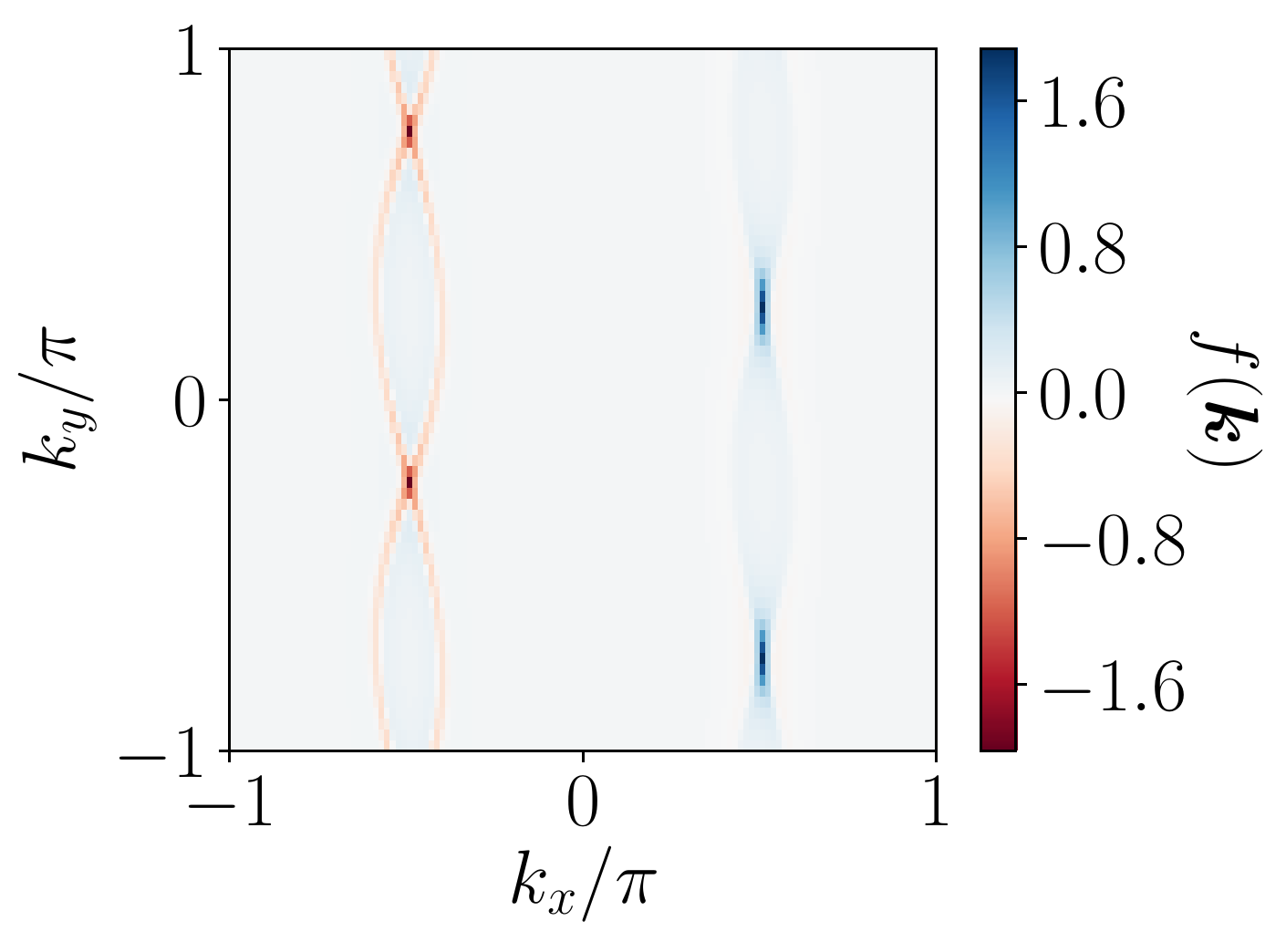}}
  \caption{Momentum dependence of real part of (a) the anomalous Green's function and (b) $f(\bm{k})$ of PDW-singlet state. 
  The center–of-mass momentum $\bm{Q}=(\pi,\pi/2)$, $t_y/t_z=t_d/t_x=0.2$, $T/t_x=0.04$, $f_s=0.97$, $N=128\times64$, $M=4096$.}
  \label{fig9}
\end{figure*}



Next, to clarify the stability of the PDW-singlet state from the viewpoint of the 
degree of the nesting of the Fermi surface, we plot the Fermi surface 
and the one that is parallelly moved by the nesting vector $\bm{Q}_{nest}=(\pi,\pi/2)$ at $t_y/t_x=t_d/t_x=0.2$ and $t_y/t_x=t_d/t_x=0.1$, as depicted in Fig.~\ref{fig8}.
The overlap of the red and blue lines near $k_x=-\pi/2$, as displayed in Fig.~\ref{fig8}, is not sufficient.
Because two electrons can form a Cooper pair only in the overlapping region of the two lines depicted in Fig.~\ref{fig8}, 
as the nesting condition becomes worse, the more PDW-singlet state is destabilized as compared with the 
superconducting states with zero center-of-mass momentum. 
To confirm this, we illustrate the momentum dependence of the real part of the anomalous Green's function of the PDW-singlet state $\mathrm{Real}[F(\bm{k},\pi T)]$ calculated under the RPA, as depicted in Fig.~\ref{fig9}(a), where the maximum value of $|\mathrm{Real}[F(\bm{k},\pi T)]|$ is normalized to be unity.
As depicted in Fig.~\ref{fig9}(a), $\mathrm{Real}[F(\bm{k},\pi T)]$ has a nonzero value only in the overlap region of the red solid and blue dashed lines in Fig.~\ref{fig8}(a), because $G(k^\prime)G(-k^\prime+Q)$ in eq.~(\ref{F}) reaches a maximum in this 
region. The effective interaction in the RPA has peaks at $\bm{Q}_{\mathrm{nest}}$ and $-\bm{Q}_{\mathrm{nest}}$, and the gap function is given by eq.~(\ref{gapf}). Therefore, we define $f(\bm{k})$ as
\begin{align}
  f(\bm{k})=F(\bm{k}+\bm{Q}_{\mathrm{nest}},\pi T)+F(\bm{k}-\bm{Q}_{\mathrm{nest}},\pi T)
\end{align}
and calculate it as depicted in Fig.~\ref{fig9}(b) to estimate the regions where the real part of the gap function of the PDW-singlet state can have large values.
We confirm that the regions where it has large values, as depicted in Fig.~\ref{fig6}(a) and \ref{fig9}(b), are almost the same.
Therefore, when the degree of nesting is lowered, the regions where the gap function of the PDW-singlet state 
can have large values are more reduced.
The nesting of the Fermi surface displayed in Fig.~\ref{fig8}(b) is more prominent than that displayed in Fig.~\ref{fig8}(a).
This means that as the one-dimensionality becomes stronger, the nesting condition becomes better. 
Therefore, the PDW-singlet state becomes stable at an extreme one-dimensional region in Fig.~\ref{fig5}.

\begin{figure}[htbp]
	\parbox{0.96\linewidth}{\centering
  \includegraphics[width=\linewidth]{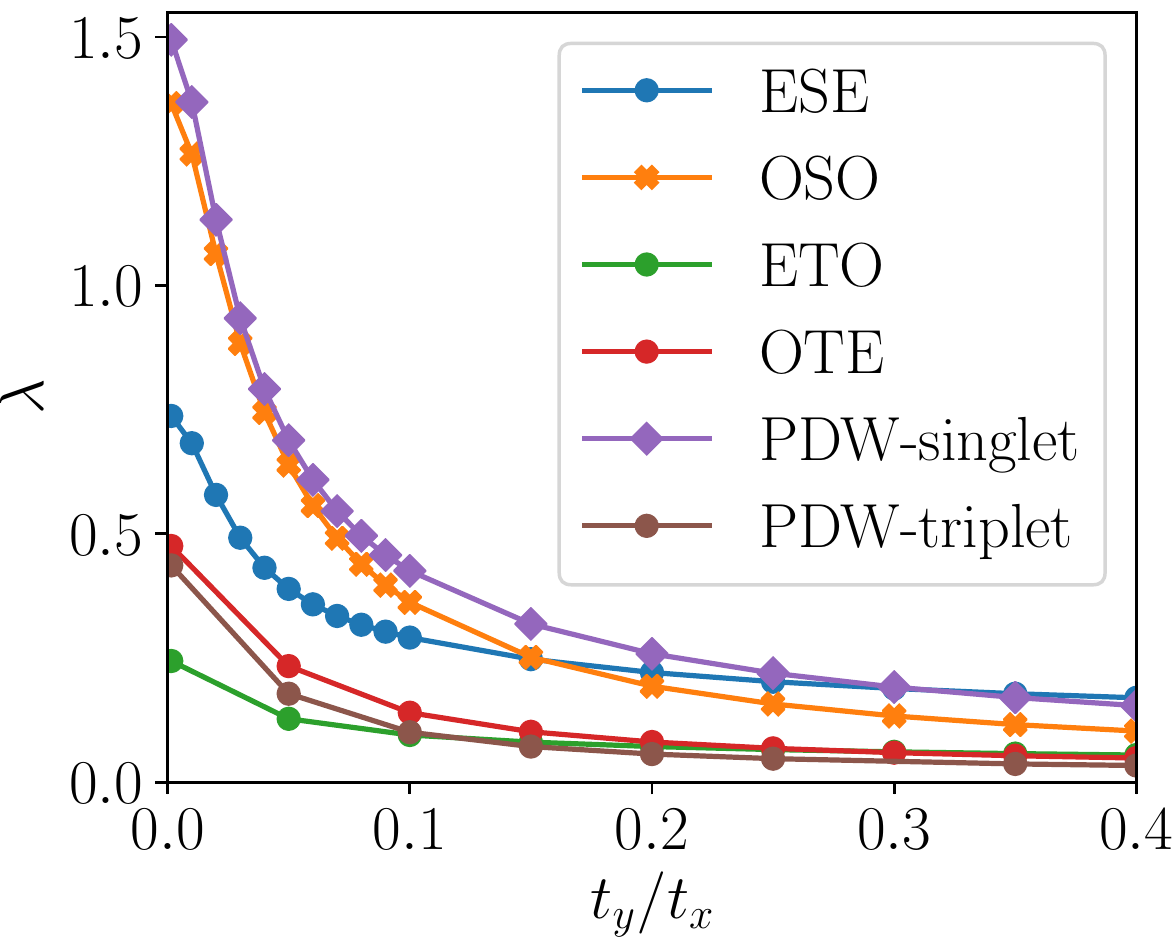}}
  \caption{$t_y/t_x$ dependence of eigenvalue $\lambda$ of the linearized Eliashberg equation calculated under RPA. The PDW-singlet state appears from $t_y/t_x=0.3$ to $0$. Center-of-mass momentum of the PDW states $\bm{Q}=(\pi,\pi)$, $t_d=0$, $T/t_x=0.04$, $f_s=0.97$, $N=128\times64$, $M=4096$.}
  \label{fig10}
\end{figure}

We fix $t_d=0$ to eliminate the instability from the incompleteness of the nesting. Therefore, the Fermi surface has the perfect nesting vector $\bm{Q}_{\mathrm{nest}}=(\pi,\pi)$.
We illustrate the $t_y/t_x$ dependence of the eigenvalue calculated using the RPA, as depicted in Fig.~\ref{fig10}.
In Fig.~\ref{fig10}, a stable region of the PDW-singlet state is expanded to $t_y/t_x\simeq0.25$ owing to the perfect nesting.
However, the ESE state is most stable in the region that is not quasi-one-dimensional.
Therefore, similar to the OSO pairing, the PDW-singlet state is stabilized because the ESE pairing becomes 
unstable in a quasi-one-dimensional parameter region \cite{Shigeta, Shigeta2}.
\par
\begin{figure}[htbp]
	\parbox{0.96\linewidth}{\centering
  \includegraphics[width=\linewidth]{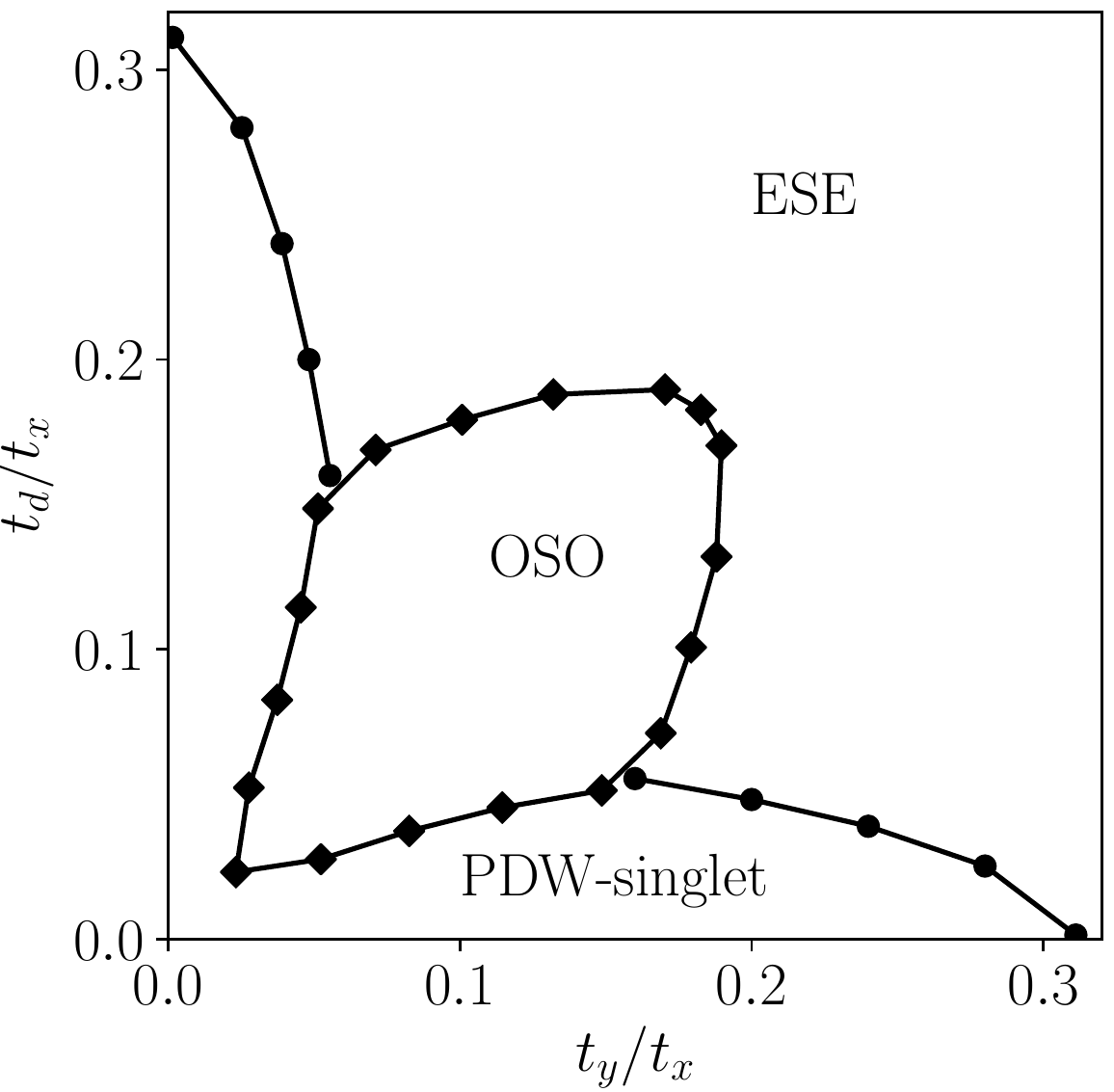}}
  \caption{Phase diagram in a quasi-one-dimensional system. We compare the maximum eigenvalue calculated by the RPA. $T/t_x=0.04$, $f_s=0.97$, $N=128\times64$, $M=4096$}
 \label{fig11}
\end{figure}

 We exhibit the phase diagram in a quasi-one-dimensional parameter region in Fig.~\ref{fig11}. 
In this calculation, we use the RPA and do not fix the on-site interaction $U$ as before.
As displayed in Fig.~\ref{fig11}, the OSO state is stabilized in the region with $t_{d} \simeq t_{y}$ 
where the spin frustration effect is prominent. In contrast, 
the PDW-singlet state becomes dominant in the region with strong one-dimensionality 
where the nesting of the Fermi surface becomes prominent. 

\par
\begin{figure}[htbp]
	\parbox{0.96\linewidth}{\centering
  \includegraphics[width=\linewidth]{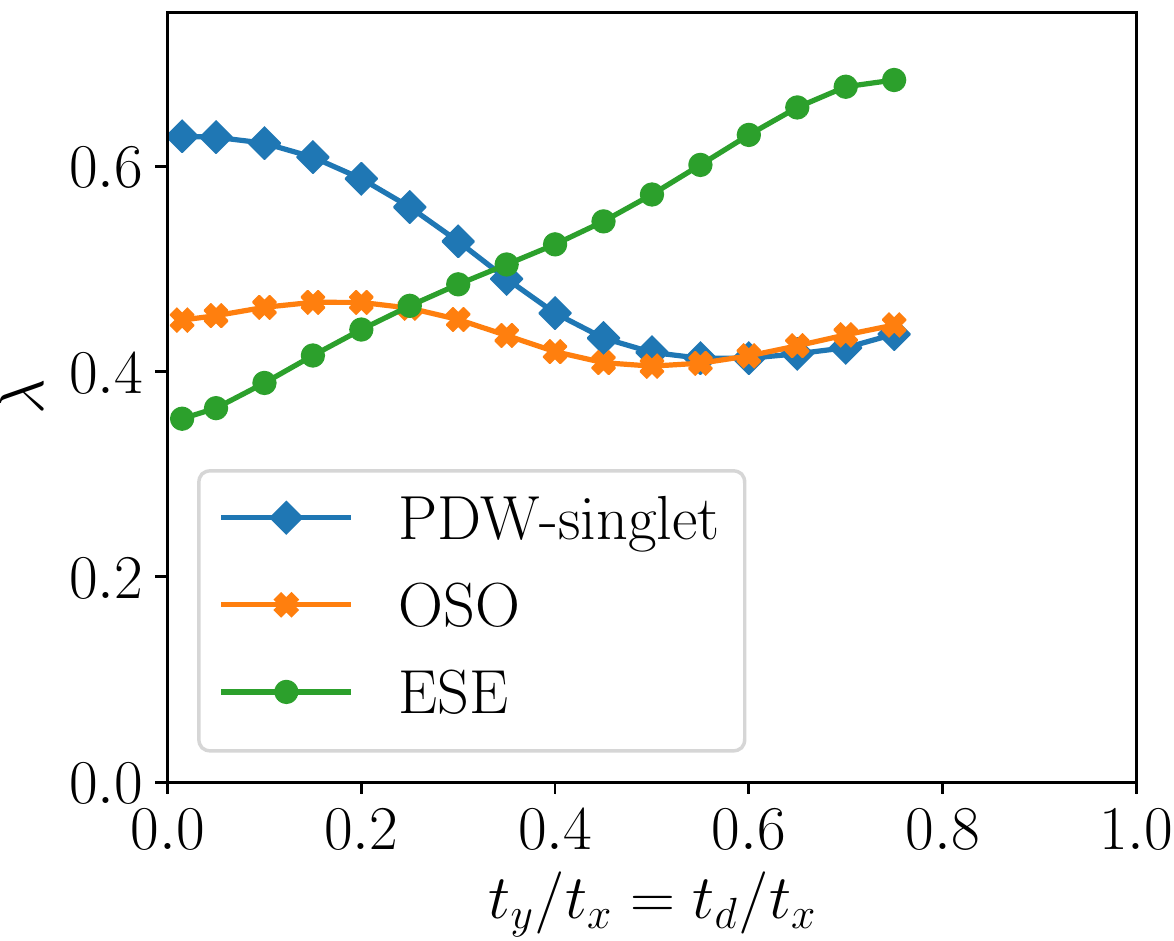}}
\caption{$t_y/t_x=t_d/t_x$ dependence of eigenvalue $\lambda$ of the linearized Eliashberg equation calculated under FLEX approximation. $T/t_x=0.04$, $f_s=0.97$, $N=128\times64$, $M=4096$.}
\label{fig12}
\end{figure}

\begin{figure}[htbp]
	\parbox{0.85\linewidth}{\centering
		\includegraphics[width=\linewidth]{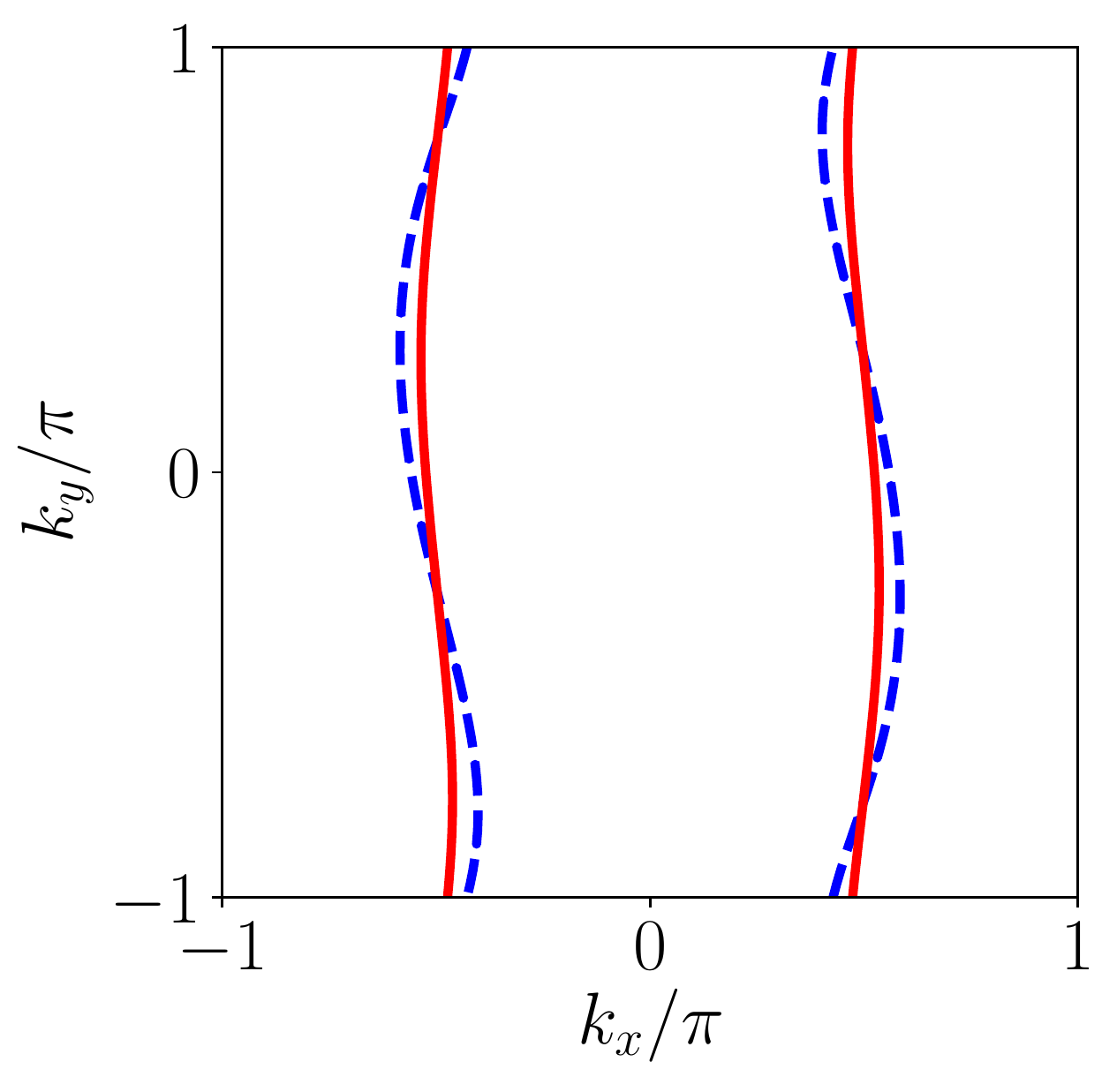}}
\caption{Fermi surface considering the self-energy. The red solid line represents the Fermi surface calculated using the FLEX approximation, and the red dashed line represents the no-interaction Fermi surface. $t_y/t_x=t_d/t_x=0.2$, $T/t_x=0.04$, $f_s=0.97$, $N=128\times64$, $M=4096$.}
\label{fig13}
\end{figure}

 Finally, we illustrate the $t_y/t_x=t_d/t_x$ dependence of the eigenvalue calculated by the FLEX approximation, as depicted in Fig.~\ref{fig12}.
We choose the nesting vector as the center-of-mass momentum of a Cooper pair, as in the case of the RPA.
In Fig.~\ref{fig12}, the PDW-singlet state is evidently stabilized in a wider parameter region as compared to the 
calculated results based on the RPA.
To understand the reason for this stabilization, we calculate the Fermi surface 
considering the self-energy obtained by the analytic continuation from 
the Matsubara frequency to the real one using the Pade approximation.
The solid red and blue lines in Fig.~\ref{fig13} represent the Fermi surfaces calculated using the FLEX approximation and without the self-energy correction, respectively. 
The line shape of the resulting Fermi surface becomes increasingly one-dimensional because of the 
self-energy. Therefore, the PDW-singlet state becomes stable owing to the improvement of the nesting.
In contrast, the OSO pairing is suppressed by the broadness of the peak width of the effective interaction by self-energy, as mentioned in previous works \cite{shigeta2013possible}. 
The peak width of the effective interaction calculated using the FLEX approximation is wider than that calculated using the RPA, 
and the PDW-singlet state does not suffer the instability, as described above, because the
principal component of the PDW-singlet state is the even-frequency state. 
For these reasons, the PDW-singlet state becomes more stable using the FLEX approximation as compared with the RPA.
On the other hand, we calculate the temperature dependence of the maximum eigenvalue calculated by FLEX at a higher temperature than $T = 0.04$. 
Then, the phase boundary between ESE and PDW-singlet state at $T=0.06$ appears at $t_y=t_d\simeq 0.5$ while it appears at $t_y=t_d\simeq 0.35$ as seen from Fig. \ref{fig12} . 
Thus, a parameter region of the PDW-singlet phase tends to decrease as the temperature decreases. 

\section{CONCLUSION}
In this study, we examined the possible pairing symmetry 
mediated by spin fluctuation 
in the quasi-one-dimensional Hubbard model 
considering the finite center-of-mass momentum of a Cooper pair, referred to as a pair density wave (PDW).
Based on the random phase approximation (RPA) calculation, 
we have compared the stability of the superconducting states such as the 
even-frequency spin-singlet even-parity (ESE), even-frequency 
spin-triplet odd-parity (ETO), and odd-frequency spin-singlet odd-parity (OSO)
pairings 
without the center-of-mass momentum and PDW with the center–of-mass momentum 2$k_{F}$
by comparing the maximal eigenvalues 
of the linearized Eliashberg equation.
Among these, the ESE $d$-wave pairing, OSO $p$-wave pairing, and 
spin-singlet PDW are the dominant competing states in the present model. 
In the quasi-one-dimensional parameter region 
$t_{y}$, $t_{d}\ll t_{x}$,  
the ESE $d$-wave pairing becomes 
unstable owing to the overlapping of the nodes of the gap function and 
the Fermi surface. 
Therefore, the obtained PDW-singlet state with the center-of-mass momentum 
$\bm{Q} \sim 2k_{F}$ 
becomes the most dominant state when one-dimensionality 
becomes prominent. 
The nodes of the present PDW state do not cross the Fermi surface, 
similar to odd-frequency spin-singlet $p$-wave pairing (OSO). 
The present PDW-singlet state is supported by the nesting of the Fermi surface and becomes unstable in the case of incomplete nesting of the Fermi 
surface because the regions where a Cooper pair can be formed are reduced. 

It should be noted that 
the obtained PDW-singlet gap function has both even-frequency and 
odd-frequency components. 
In general, the obtained gap function of the PDW-singlet state is neither an 
odd nor an even function 
owing to the inversion of the momentum ${\bm k}$ to ${-\bm k}$. 
In the present case, because the 
magnitude of the center-of-mass momentum is almost 2$k_{F}$, 
the even-frequency part has a momentum dependence similar to that of the OSO component.
The even-frequency component of the gap function is dominant, and this node does not overlap with the Fermi surface. 
%
%
\par 
Under the FLEX approximation, 
the parameter region where PDW is stabilized becomes wider 
as compared to that of RPA because the 
OSO state becomes suppressed owing to the self-energy effect of the quasiparticle.

In this study, we have only considered the on-site Coulomb repulsion. 
It is known that charge fluctuation is enhanced in the presence of the off-site 
Coulomb repulsion, which also enhances the spin-triplet $f$-wave 
pairing in the Q1d superconductor~\cite{Kuroki2004}. 
The charge fluctuation can also enhance odd-frequency 
spin-triplet pairing in Q1d superconductors \cite{Shigeta2}. 
It is interesting to see how the PDW triplet state is stabilized in the presence of the off-site Coulomb repulsion.

The physical properties of the PDW-singlet state are also interesting. Because this pairing has a spatial oscillation, the translational invariance is broken microscopically, and we expect a sufficient odd-frequency pair amplitude even if the major component of the gap function is even-frequency pairing \cite{odd1,odd3,odd3b,Eschrig2007,tanaka12}. 
Furthermore, it is known that odd frequency pair density wave is induced by the coexsistence of the charge density wave and d-wave superconductor in underdoped cuprates \cite{Chakraborty2021}.
It is also interesting to study the
tunneling, Josephson, and proximity effects in this pairing \cite{TK95,kashiwaya00}. 
\\~\\
{\bf ACKNOWLEDGMENTS}
\\~\\
This work is supprted by 
Grant-in-Aid for Scientific Research B (KAKENHI Grant No. JP18H01176 and No.
JP20H01857),  Research A (KAKENHI Grant No. JP20H00131)
and the JSPS Core-to-Core program Oxide Superspin International
Network.

\bibliographystyle{apsrev4-1}
\bibliography{paper3}

\end{document}